\documentclass[floatfix, preprintnumbers, amsmath, amssymb, 10pt]{revtex4}
\usepackage{graphicx}
\usepackage{dcolumn}
\usepackage{bm}
\usepackage{epsfig}

\begin{document}
\title{D0 Matrix Mechanics and Topology Change of Fuzzy Spaces}

\author{Subodh P Patil}
\email[Electronic address: ]{patil@het.brown.edu}

\affiliation{Dept.of Physics, Brown University,\\
Providence R.I. 02912, U.S.A.} 

\date{July 21 2004}
\revised{\today}

\begin{abstract}

We consider the physics of a matrix model describing D0-brane dynamics in the presence of an RR flux background. Non-commuting spaces arise as generic soltions to this matrix model, among which fuzzy spheres have been studied extensively as static solutions at finite N. The existence of topologicaly distinct static configurations suggests the possibility of D-brane topology change within this model, however a dynamical solution interpolating between topologies is still somewhat elusive. In this paper, we study this model in the limit of infinite dimensional matrices, where new solutions-- the fuzzy cylinder and the fuzzy plane among them-- appear. We argue that any dynamics which involves topology change will likely only occur in this limit, after which we study the decay of a fuzzy cylinder into an infinite collection of fuzzy spheres as both a classical and a quantum phenomenon. We conclude from this excercise that in certain limits, matrix models offer a viable framework in which to study topological dynamics of fuzzy spaces, and could perhaps be a precursor to a viable theory of space-time topological dynamics.       

\end{abstract}

\maketitle

\section{Introduction and Motivation}

Non commutative geometries have long been studied extensively in their own right as a possible framework for spacetime physics at very high energy scales\footnote[1]{See \cite{k1}\cite{k2}\cite{k3} for reviews on the subject}. From their genesis as an ad-hoc manner in which to attempt a regularization of quantum field theory \cite{k6}, non commutative geometries have since been motivated by semi-classical studies in general relativity\cite{k8} and string theory. As an area of intrinsic mathematical interest\cite{k1} it apparently needs no justification. More recently such geometries have found a natural realization as solutions of various matrix models describing D-brane dynamics. (see \cite{k4}\cite{k5}\cite{pol} and refernces therein). It is the aim of this report to study a particular matrix model describing D0-brane configurations in the presence of RR flux as a concrete example in which one finds interesting time dependent physics for these non commutative spaces, including possible topology changing transitions between distinct geometries. By topology in the context of non commutative spaces, we specifically refer to the properties of the commutative manifold associated with the fuzzy geometry we are considering. Hence by topology changing dynamics, we are refering to dynamics which in the commutative limit (if such a limit is available), would look like topology change in the classical sense. It appears that this putative aspect of matrix dynamics is closely related to its nature as a theory with $U(N)$ invariance. The eventual hope is that matrix models could provide a realistic framework in which to study the toplogical dynamics of extended objects embedded in spacetime, if not spacetime itself-- a widely anticipated (if sometimes controversial) corollary of any putative quantum theory of gravity, and one that is deserving of a brief review.\linebreak

\par

Ever since Wheeler's initial musings on the matter \cite{k7}, the notion of fluctuating spacetime topology lay more or less dormant in the minds of theoretical physicists for several decades. Although most physicists have long agreed that notions of spacetime would have to be radically reformulated at the Planck scale, we have only just begun to uncover the physics likely to be responsible for this reformulation. Inspite the multitude of approaches to modelling spacetime at the Planck scale (among them, canonical and covariant quantum gravity, loop quantum gravity, string theory and non-commutative geometry), common to them all is the notion of a minimally resolvable length scale. This notion is either taken up as an a priori input to the theory, or is a derived consequence of other principles. For instance, in loop quantum gravity, it is a non-trivial result of this theory that various geometric operators such as volume and area are quantized in units of $\hbar$\cite{rov}\footnote{As for canonical quantum gravity, the metric tensor and its momentum conjugate with respect to the Einstein- Hilbert action are taken to be a (constrained) canonical pair. Although strictly speaking spacetime is still taken to be commuting, the metric tensor and its momentum conjugate satisfy Heisenberg inequalities, which effectively determines an energy dependent minimal length scale. The formation of singularities above a certain energy scale imply that there is an absolute minimum length scale in this theory. Slightly different considerations in semi-classical general relativity, can be shown to derive Heisenberg inequalities for spatial coordinates\cite{k8}, which imply a non-commuting structure to spacetime and hence imply non-commutative geometry.}. In string theory, T-duality, the superstring uncertainty principle and the appearance of non-commutative geometry (NCG) in certain situations all imply a minimal length scale, though its interpretation depends on precisely which of the three mechanisms are at work\footnote{See \cite{paddy} for a demonstration of how T-duality as an a priori assumption in and of itself implies the appearance of a regularizing minimal length in field theoretical amplitudes}. In NCG, one takes as a starting point the non-commuting nature of space (and/or time), which is logically equivalent to positing a spacetime uncertainty principle and formulates physics afresh in this framework. Non-commutative geometry naturally makes an appearance in string theory through the dynamics of D-branes \cite{k4}\cite{k5}\cite{k12}\cite{pol} and through the theory of relativistic membranes \cite{wati}. It will almost certainly have a central role in the future if the `M' in M-theory turns out to stand for matrix. Hence from the perspective of a string theorist, NCG would be the  most natural and best motivated avenue in which to directly study the nature of spacetime at the Planck length.\linebreak 

\par

After the hard work of consistently formulating any putative model of spacetime at the Planck length has been done, the question of whether or not topology changing dynamics is a feature of this model, or whether the theory breaks down before such dynamics can take place, immediately beckons. In considering this question, perhaps it is appropriate to refresh ourselves with what the presently accepted paradigm of space time dynamics (classical general relativity) has to say. It should be clear that general relativity (derived from the Einstein-Hilbert action), does not contain any scope for topological dynamics. One can do general relativity in a background of any given topology, but the resulting dynamics does nothing to change this topology. This is easily seen from the fact that constructively, any topology changing process will involve the formation of singularities, which is where general relativity breaks down. Furthermore, it has been demonstrated \cite{witt} that general relativity excercises a sort of `topological censorship' principle, in that at intermediate energies (below the compactification scale and above the Hubble scale) all causal curves (probes) that traverse non-trivial toplogical features of spacetime causally disconnect from the causal curves (observers) which do not. That is, the most we will ever observe of a non-trivial topological feature of spacetime is a singularity.\linebreak 

\par
Adopting a `top down' approach, that is one which bases itself on established physical paradigms, euclidean quantum gravity posits a sum over histories approach to quantizing general relativity. Topology changing processes have been extensively studied in this context (see \cite{fay} for a review and references to the relevant literature). By this we mean that within the sum over histories, histories which interpolate between 3-d manifolds which are non-diffeomorphic to each other (cobordisms) are analysed for their likely contribution to the path integral. This program has had some degree of success, particularly in ruling out certain classes of acausal topological transitions (cobordisms)\cite{fay} within this framework. However the approach still has a lot of fundamental issues to answer for, not least of which being the definition and consistency of the sum of histories (path integral) approach to quantizing general relativity.\linebreak 

\par

String theory on the other hand, has already conclusively demonstrated the capability to model spacetime topology change (\cite{bg1}\cite{bg2} and references therein). Starting from the observation that the K\"ahler moduli space (which defines the Calabi-Yau 3-fold on which the theory is compactified, and hence the CFT describing it on the worldsheet) separates into distinct cells which define distinct compactifications, Aspinwall, Greene and Morrison\cite{bg2} showed that there was no physical obstruction to marginally deforming the underlying worldsheet CFT such that it moves into a different cell in moduli space. This has the effect of changing the target space of the theory in such a way that in certain cases, the topology of the target space has changed. Such geometrical transitions are continuous and are known as 'flop' transitions (previously known to algebraic geometers), and have been extensively studied since. However it is in keeping with the perspective natural to string theory-- that geometry is notion derived from the dynamics of our degrees of freedom-- that the study of such transitions involves the geometry only indirectly. The calculations involved are quite formidable, and although this remarkable result is not to be detracted from, if our primary concern is to directly model quantum spacetime dynamics, one would be tempted to look for another avenue to explore this question. This temptation only becomes stronger when one would like model more than just the dynamics of the internal space of a given compactification.\linebreak

\par

Non-commutative geometry seems to offer exactly such an avenue, and as delineated by Madore and Saeger\cite{madore} should rather naturally describe topology changing dynamics, once the dynamics have been determined. Matrix models (in a certain reference frame) naturally offer just such a dynamical framework. We stress that the particular framework that we are about to work in, will specifically concern examining D-brane topology change. We will be studying the dynamics of non commuting spaces which appear as solutions to a rather well known matrix model describing D0-brane dynamics, with the aim of demonstrating the existence of dynamical topology change in this context. One only has to realise the possibility that our universe might be a 3-brane embedded in a higher dimensional spacetime to appreciate the relevance of this study towards the eventual goal of understanding spacetime topology change. 

\par

Henceforth, any refernce to topological dynamics will specifically refer to dynamics in the context of non-commutative geometry. However even within this domain, there are different approaches in the literature concerning the type of topological transition studied. We wish to point out that to our knowledge, the type of non-commutative topological transition we wish to report on is dictinct from the types hitherto looked at (some of which have demonstrated promising preliminary results-- see for example \cite{k4}\cite{k9}\cite{patta}\cite{topos1}). To understand this, it helps to first categorize the types of possible topological transitions thought possible in the context of non-commutative geometry. 

\subsection{Classes of Topological Transitions}

We start with the basic observation that matrix models often have as solutions to their (static) equations of motion, matrices that define an algebra. In the event that this algebra is a Lie algebra, they satisfy the relation:

\begin{equation}
\label{alg}
[X_i,X_j] = if_{ijk}X_k
\end{equation}   

That is, any set of matrices that satisfies the above is a solution to the matrix model in question. The structure constants $f_{ijk}$ define the Lie algebra. In general, for any given dimension N of the matrices in question, there are many solutions to (\ref{alg}) which correspond to the number of ways N can be partitioned into the integers defining the irreducible representations of the Lie algebra. That is, one could have any direct sum of reducible representations consistent with the total dimension of the matrices being equal to N. Of course, one also has the irreducible solution (if such an N dimension irrep exists). Furthermore, if (\ref{alg}) also describes the structure relations of a non-commutative geometry, one can interpret this as saying that the matrix model has solutions corresponding to multiple copies of the fuzzy spaces defined. To illustrate this, let us hypothetically take the index $i$ to run from 1 to 3, and the $f_{ijk}$ to be propotional to the 3-d Levi-Civita tensor:

\begin{equation}
\label{su2}
[X_i,X_j] = i\kappa\epsilon_{ijk}X_k
\end{equation}

In which one recognizes the $X_i$ to define the structure relations of the fuzzy sphere (see \cite{k10} for a review of its construction), with the non-commutativity parametrized by $\kappa$. Any representation of su(2) (irreducible or otherwise) given by the generators $\{J_i\}$ furnishes the solution:

\begin{equation}
\label{su2.1}
X_i = \kappa J_i
\end{equation} 

For an irrep, the casimir operator $X_iX_i = \kappa^2j(j+1)$ gives us the square of the radius of the fuzzy two sphere, where the dimension of the representation is given by $N= 2j + 1$. Any reducible representation can be expressed as a direct sum of irreps:

\begin{equation}
\label{rep}
X_i = \oplus_{r} \kappa J_i^{(r)}
\end{equation}

Where the summation $r$ runs over all the irreps contained in the representation. This solution descirbes a set of multiply superimposed spheres. The radius of each constituent 2-sphere is given by $R^2_r = 
\kappa^2 j_r(j_r+1)$ and the dimension of the representation is now given by $N 
= \sum_r (2j_r+1)$.\linebreak  

\par

Although the equations of motion do not fix the representation, the energy of the solutions corresponding to different representations will differ in general. The example we are presently studying can be derived from the matrix model defined by the action:

\begin{equation}
\label{act}
S = T_0 \int dt Tr \Bigl( \frac{1}{2} \dot{X}_i \dot{X}_i + 
\frac{1}{4}[X_i,X_j][X_i,X_j] - \frac{i}{3}\kappa \epsilon_{ijk}X_i[X_j,X_k] 
\Bigr)
\end{equation}

We will introduce this model more formally in the next section. The energy of the static solutions is given by:

\begin{equation}
\label{V}
V = Tr \Bigl( -\frac{1}{4}[X_i,X_j][X_i,X_j] + \frac{i}{3}\kappa 
\epsilon_{ijk}X_i[X_j,X_k] \Bigr)
\end{equation}

From which we see that the energies of our fuzzy sphere solutions are given by:      

\begin{equation}
\label{esph}
E = -T_0\kappa^4\frac{1}{6}\sum_r j_r(j_r+1)(2j_r+1)
\end{equation}

Where again the summation is over the irreps contained in the solution. We can immediately conclude that for a given N, the irreducible representations correspond to the lowest energy (bound state) solutions. The observation that reducible representations have a higher energy has motivated the study of transitions between these representations. Geometrically, the simplest example of this would correspond to two spheres melding into one (\cite{patta}\cite{k4} and references therein). This would certainly involve a change in topology, as two disconnected spaces fuse into one connected space. We shall refer to this type of topological transition as `{\it isogeometric}'. That is the algebraic and geometric structure (\ref{su2}) of the solutions does not change in the transition.\linebreak    

\par

A second class of topological transitions was studied in \cite{k9}, where the observation that a given Lie algebra structure could give rise to several distinct fuzzy geometries was used to study possible transitions between these geometries. The example considered was the case when (\ref{alg}) defined the Lie algebra of $SU(4)$, which contains fuzzy versions of $CP(3)$, $CP(2)$ and $S^2\times S^2$, among others \cite{k9}. Transitions between these fuzzy geometries were explored and obstructions to topology change were uncovered arising from the equations of motion which fixed the particular representation of $SU(4)$. Nevertheless, this work does highlight the possibility of a second type of topological transition, which we shall classify as `{\it isoalgebraic}'. That is, that although the geometry and topology of the fuzzy spaces might undergo a transformation (e.g. $CP(2) \to S^2\times S^2$), the algebraic structure of the matrix solutions (\ref{alg}) is preserved in this transition. Although an isogeometric transition is also isoalgebraic, the converse need not be true.\linebreak

\par

In the present work, we wish to study a third class of topological transitions which are neither isogeometric nor isoalgebraic. By this, we mean a transition such that the algebraic structure of the solutions changes in time, with the initial algebra describing one fuzzy geometry and the final algebra, describing another. The specific example we will be studying will be the decay of a fuzzy cylinder into an infinite collection of fuzzy spheres. This study is motivated by the observation that a rather well known matrix model which we are about to introduce, contains solutions corresponding to different algebras in the limit of infinite dimensional matrices\cite{sub}. This limit is neccesitated by the fact that some of the solutions of this model only have infinite dimensional representations. Although going to such a limit might seemingly involve a loss of calculational ease, for the questions we are interested in, we will find that this is not the case. In fact we feel that there is a case for why one is likely to encounter toplogical dynamics only in this limit, and in a sense this limit could be a prerequisite for our investigation of topology change.

\subsection{Topology change only in the Large N Limit?}

A generic feature of matrix models of interest, is their global $U(N)$ invariance. The action defining our example (\ref{act}) is invariant under the symmetry $X_i \to UX_iU^\dag$, and it is this feature alone of matrix actions that has generated much theoretical interest. It has been observed that as $N$ tends to infinity, the symmetry group $U(N)$ tends to the area preserving diffeomorphism groups of certain 2-dimensional surfaces, such as $S^2$ or $T^2$. This association arises in the matrix regularization of membrane theory, where the original diffeomorphism invariance of the worldvolume manifests as a $U(N)$ invariance of the associated matrix model. However, it turns out that one ends up with the same $U(N)$ invariance when one regularizes membranes of any given genus\cite{wati}. This is due to the fact that the group of area preserving diffeomorphisms of any 2-dimensional surface (of arbitrary genus) can be approximated by $U(N)$ in the large $N$ limit \footnote{See \cite{wati} and references therein-- \cite{crazy} in particular. For a less mathematically cumbersome treatment, which arrives at the conclusion that $U(N)$ as $N \to \infty$ should contain as subgroups, the diffeomophism groups of many 2-dimensional surfaces, see \cite{js}}. That is, the matrix regularized theory appears to have more structure than the theory it was meant to approximate by this virtue of $U(N)$, and seems to describe a theory where the membrane topology can change. From this observation we feel that it is highly plausible that any matrix model with $U(N)$ invariance could, and should only describe the topological dynamics of fuzzy spaces in this large $N$ limit. This observation seems to have borne out in practice as the example we are about to study, (which involves infinite dimensional matrices) has proven to be remarkably tractable when compared to studying similar questions in a finite dimensional setting.

\section{The Model}

The matrix model alluded to previously, which we will be studying is described by the following action, describing D0 brane physics in the presence of an RR flux background (\cite{k4}\cite{k5}\cite{k9}\cite{patta}\cite{sub}\cite{cs1}):

\begin{equation}
\label{act2}
S = T_0 \int dt Tr \Bigl( \frac{1}{2} \dot{X}_i \dot{X}_i + 
\frac{1}{4}[X_i,X_j][X_i,X_j] - \frac{i}{3}\kappa \epsilon_{ijk}X_i[X_j,X_k] 
\Bigr)
\end{equation}

We follow the conventions of \cite{k4} and work in units where $2\pi\alpha' 
= 1$. The $X_i$ are taken to be $N \times N$, hermitian matrices which can be taken to be traceless, and the D0 brane tension is given by $T_0 = \sqrt{2\pi}/g_s$. The index $i$ runs from 1 to 3. This action is supplemented by the Gauss law constraint:

\begin{equation}
\label{Gauss}
[\dot{X}_i,X_i] = 0
\end{equation}

Which arises from the $A_0$ equation of motion for the RR gauge field. Note that the last term in the action is a Chern-Simons term which induces interactions between the D0-branes through the 4-form flux, which assumes the vacuum expectation value $F^{4}_{0123} = -2\kappa\epsilon_{ijk}$. This term was deduced by Myers by demanding consistency of the D-brane action with T-duality \cite{k5}. This model also appears in several other contexts outside of D-brane physics (see \cite{k4} and references therein). However we shall adopt the perspective of a non-commutative geometer in this report, and only utilise this model insofar as it provides a model for the matrix dynamics of fuzzy spaces. The equations of motion derived from this action are:

\begin{equation}
\label{seom}
\ddot{X}_i = [X_j, \bigl([X_i,X_j] - i\kappa \epsilon_{ijk}X_k \bigr)]
\end{equation}

Inspite of its appearance, the above results whether or not one takes the Gauss law constraint into account (the only modification would be terms proportional to the constraint itself, which vanish by the equations of motion for the auxilliary variable if the constraint is formulated appropriately)\cite{sub}. From the above, we can immediately read off two classes of static solutions-- commuting matrices,and fuzzy two spheres:

\begin{equation}
\label{fuzzysp}
[X_i,X_j] = i\kappa \epsilon_{ijk}X_k
\end{equation}

Where in all but name, various solutions and their energetics were discussed in the introduction (\ref{su2.1}) - (\ref{esph}). From the form of (\ref{seom}), one might think that we've exhausted the set of possible static solutions, however a slight recasting of the problem will show that this is not the case. Let us perform the following change of variable:

\begin{equation}
\label{raise}
X_+ = X_1 + iX_2 ~,~ X_- = X_1 - iX_2~;~ X_+^\dag = X_-
\end{equation}

After which, the equations of motion become:

\begin{equation}
\label{lc1}
\ddot{X}_+ ~= ~\frac{1}{2}[X_+, [X_+,X_-] ] ~+ ~[X_3, [X_+,X_3] + 2\kappa 
X_+ ]
\end{equation}

\begin{equation}
\label{lc2}
\ddot{X}_- ~= ~\frac{1}{2}[X_-, [X_-,X_+] ] ~+ ~[X_3, [X_-,X_3] - 2\kappa 
X_- ]
\end{equation}

\begin{equation}
\label{lc3}
\ddot{X}_3 ~= ~\frac{1}{2}[X_+, [X_3,X_-] ] ~+ ~\frac{1}{2}[X_-, [X_3,X_+] ] 
~+ ~\kappa [X_+,X_-]
\end{equation}

And the Gauss law condition becomes:

\begin{equation}
\label{gc}
[\dot{X}_3,X_3] ~+ ~\frac{1}{2}[\dot{X}_+,X_-] ~+ 
~\frac{1}{2}[X_+,\dot{X}_-] ~= ~0
\end{equation}

Equations (\ref{lc1}) and (\ref{lc2}) are adjoints of each other, and so we are left with only two independent matrix equations to satisfy. The fact that the equation governing either $X_+$ or $X_-$ has in general a hermitian and an anti-hermitian part accounts for the two hermitian equations of motion that we had previously. We now try the ansatz:
 
\begin{equation}
\label{ansatz}
[X_+,X_-] = \lambda X_3 ~,~ [X_3,X_\pm] = \pm \theta X_\pm
\end{equation}

For which eqs. (\ref{lc1})-(\ref{lc3}) imply

\begin{eqnarray}
\label{cond}
\ddot{X}_+ &=& (2\kappa - \theta - \lambda /2)\theta X_+\\ \ddot{X}_3 &=& 
(\kappa - \theta)\lambda X_3
\end{eqnarray}

Looking for static solutions, we immediatly spot three distinct classes of solutions-- commuting matrices ($\lambda = \theta = 0$), fuzzy spheres ($\lambda = 2\kappa$, $\theta = \kappa$) and something else corresponding to $\lambda = 0$, $\theta = 2\kappa$. That we identify the solution $\lambda = 2\kappa$, $\theta = \kappa$ with the fuzzy sphere is readily seen:

\begin{equation}
\label{su2rl}
[X_+,X_-] = 2\kappa X_3 ~,~ [X_3,X_\pm] = \pm \kappa X_\pm
\end{equation}

Which is none other than the algebra of $su(2)$ in the Cartan-Weyl basis. It should be clear that (\ref{raise}) constructs the usual raising and lowering operators if the $X_i$ are taken to be elements of $su(2)$. The third solution we uncovered corresponding to $\lambda = 0$, $\theta = 2\kappa$, happens to describe the fuzzy cylinder\cite{sub} (see \cite{k11} a review of its construction): 

\begin{equation}
\label{cyl}
[X_+,X_-] = 0 ~,~ [X_3,X_\pm] = \pm 2\kappa X_\pm
\end{equation}

Very briefly, we outline its construction since it is not as commonly known as the fuzzy sphere. Consider the commutative cylinder, parametrized by coordinates $\tau ~ \epsilon ~ \mathbb R$ and $\phi ~ \epsilon ~ [0,2\pi]$. Functions on the cylinder can be expanded in terms of the following basis:

\begin{equation}
\label{basis}
\{ e^{in\phi} \}_{n \epsilon \mathbb Z} ~;~ f(\tau,\phi) = \sum_n c_n(\tau) 
e^{in\phi}
\end{equation} 

Define $x_+ = \rho e^{i\phi}$ and $x_- = \rho e^{-i\phi}$, where $\rho$ is 
the radius of the cylinder. All smooth functions on the cylinder can be expressed in terms of a power series in these variables. Furthermore, we have the relation $x_+x_- = \rho^2$. Being a symplectic manifold, the cylinder naturally admits the Poisson brackets:

\begin{equation}
\label{pb}
\{f,g\} := \frac{\partial f}{\partial \tau}\frac{\partial g}{\partial \phi} 
- \frac{\partial g}{\partial \tau}\frac{\partial f}{\partial \phi}
\end{equation}

where $f$ and $g$ are arbitrary functions on the cylinder. The Leibniz 
rule implies that all Poisson brackets can be generated from the following elementary brackets:

\begin{equation}
\label{fund}
\{\tau,x_\pm\} = \pm i x_\pm ~,~ \{x_+,x_-\} = 0
\end{equation}

The relation $x_+x_- = \rho^2$ is now a central element of the Lie algebra of smooth functions on the cylinder (with the Poisson brackets assuming the role of the Lie bracket):

\begin{equation}
\label{cent}
\{\tau,x_+x_-\} = \{x_\pm,x_+x_-\} = 0
\end{equation}

Derivatives of functions on the cylinder can now be effected by the action 
of the brackets:

\begin{equation}
\label{derv}
\partial^2_\phi f = \{\tau,\{\tau,f\}\} ~;~ \partial^2_\tau f = 
\frac{1}{\rho^2}\{x_-,\{x_+,f\}\}
\end{equation}

One obtains the non-commutative cylinder by `quantizing' the Poisson structure of the commutative cylinder. This is effected through the prescription $\{ , \} \to \frac{i}{\lambda}[ , ]$ where $\lambda$ is the non-commutativity parameter. Although one could convince themselves of the sense in this prescription through a number of heuersitic arguments, this is in fact a very well defined and well motivated prescription for constructing non commuting geometries \footnote{See \cite{k12} and references therein for a lucid introduction to the quantization of classical manifolds.}. With this in mind, we can write down the structure relations for the fuzzy cylinder as:

\begin{equation}
\label{fuzzyc}
[\tau,x_\pm] = \pm \lambda x_\pm ~,~ [x_+,x_-] = 0 ~,~ x_+x_- = \rho^2
\end{equation}    

Thus taking $\lambda = 2\kappa$, we see that (\ref{cyl}) is in fact the fuzzy cylinder if we make the identifications $X_3 = \tau$ and $X_\pm = x_\pm$. Note that the equations of motion (\ref{cyl}) do not impose any relations among the $X_i$ other than the algebra (\ref{cyl}). The restriction $X_+X_- = \rho^2$ is something we impose in order to make an identification with the fuzzy cylinder. From the relation $[X_3,X_\pm] = \pm 2\kappa X_\pm$, one can immediately deduce that the spectrum of $X_3$ is integer multiples of $2\kappa$, and the action of $X_+$ in the basis where $X_3$ is diagonal is that of $\rho$ times the elementary shift operator:

\begin{equation}
\label{shift}
X_3|n\rangle = 2\kappa n |n\rangle ~\to~ X_+|n\rangle = \rho |n+1\rangle
\end{equation}

($X_-$ performs the opposite shift). However we have just uncovered a very important issue concerning this solution, namely that any representation of the cylinder algebra will necessarily be infinite dimensional. This should be apparent from (\ref{fuzzyc}), which does not describe a semi-simple Lie algebra (re-expressing the algebra in terms of $X_1$ and $X_2$, we see that they form a proper ideal in the algebra). Hence any unitary representation (i.e. one in which the generators $X_i$ are hermetian-- a structural requirement of this model) will necessarily be infinite dimensional. There were two crucial aspects of the trace which we used in order to arrive at the equations of motion (\ref{seom}), namely the positive definiteness of the trace norm, and the cyclicity of the trace. The first property was used to deduce the equations of motion from the statement that the action is extremized by all variations in the fields:

\begin{equation}
\label{trnorm}
Tr \Bigl( \frac{\delta L}{\delta X_i} \delta X_i \Bigr) = 0 ~\forall~ \delta 
X_i ~\to~ \frac{\delta L}{\delta X_i} = 0
\end{equation}    

And the second property was used to compensate for the non-commutativity of the $\delta X_i$ with the $X_i$ to bring all the variations to one side of the trace. It is this property that fails in general for infinite dimensional matrices (c.f. $Tr [x,p] \neq 0$ where $x$ and $p$ are a canonical pair). However one can show that if taken carefuly, the equations of motion (\ref{seom}) remain unchanged in the limit of infnite dimensional matrices, except for one important respect-- one finds that the model now admits a `central extension' wherein the equations of motion become \cite{sub}:

\begin{equation}
\label{ceom}
\ddot{X}_i = [X_j, \bigl([X_i,X_j] - i\kappa \epsilon_{ijk}X_k \bigr)] + c_i I
\end{equation}   

One finds that this central extension allows for solutions corresponding to the fuzzy plane and the warped fuzzy plane, but this shall not concern us here. We are interested in the case where the $c_i$ all vanish which permits the solutions that we have hitherto derived: the fuzzy sphere and the fuzzy cylinder.\linebreak

\par

One should note that if the fuzzy cylinder and the fuzzy sphere solutions are to be meaningfully compared (or in the situation we're interested in, dynamically interpolated), they have to be represented in the same space (in our case, the usual Hilbert space $\mathcal H$ of square integrable functions on $\mathbb C$ \cite{sub}\cite{k11}). However it should be clear that $su(2)$ does not posses any infinite dimensional unitary irreducible representations. That is any infinite dimensional representation of $su(2)$ is necessarily reducible, which means that it is going to contain an infinite number of finite dimensional fuzzy spheres. With this in mind, let us compare the energy densities of the fuzzy cylinder solution with an infinite collection of identical spin j fuzzy spheres. One finds after evaluating (\ref{V}), the energy density $\epsilon$ ($= E/TrI$) of the fuzzy cylinder is given by:

\begin{equation}
\label{epcyl}
\epsilon_{cyl} = \frac{2}{3}\kappa^2\rho^2 T_0
\end{equation}   

And the energy density of the infinite collection of spin j spheres is given by:

\begin{equation}
\label{epsph}
\epsilon_{sph} = -\frac{1}{6}\kappa^4 j(j+1) T_0
\end{equation}

Furthermore, setting the radius of the cylinder to be equal to the radius of any one of the fuzzy spheres ($\rho^2 = \kappa^2j(j+1)$) we see that the two energy densities are given by:

\begin{equation}
\label{compa}
\epsilon_{cyl} = \frac{2}{3}\mu ~,~\epsilon_{sph} = -\frac{1}{6}\mu~~;~~ \mu = \kappa^4 j(j+1)T_0 
\end{equation} 

Which strongly suggests the possibility of a transition between these two solutions, provided the energy landscape of this model does not obstruct the transition in any way. In \cite{k4} and \cite{patta} the energy landscape of finite dimensional solutions were carefully studied and transitions between various combinations of fuzzy spheres suggested by the energetics of the model were explored, with some promising results. Although the study of the energy landscape of this model in the case of infinite dimensional matrices is quite a formidable one, we find that a few careful considerations suffice to pin down an ansatz which is an ideal candidate to model the decay of the fuzzy cylinder solution into a collection of an infinite number of spin j fuzzy spheres.

\section{The Ansatz}

As a warm up, let us consider a transition from the cylinder to a fictitious infinite dimensional irrep of the fuzzy sphere. We begin in this way as the treatment is relatively straight forward, and the techniques and results we uncover here carry over easily to a more realistic setting-- the study of the decay of the cylinder into an infinite collection of spheres. According to the equations of motion (\ref{lc1})-(\ref{lc3}), the two static solutions which our matrix model admits as solutions in the limit of infinite dimensional matrices are:

\begin{equation}
\label{s}
[X_+,X_-] = 2\kappa X_3 ~,~ [X_3,X_\pm] = \pm\kappa X_\pm
\end{equation}

which describes the fuzzy 2-sphere solution, and:

\begin{equation}
\label{c}
[X_+,X_-] = 0 ~,~ [X_3,X_\pm] = \pm 2\kappa X_\pm ~,~ X_+X_- = \rho^2
\end{equation}  

which describes the fuzzy cylinder. In the basis where $X_3$ is taken to be diagonal, we see that for an irreducible representation of the sphere, $X_3$ has eigenvalues $\kappa n$ where $n$ is an integer, whereas for the cylinder, $X_3$ has eigenvalues $2 \kappa n$ (we focus on the integer representations of the sphere for simplicity). Because of the relation $X_+X_- = \rho^2$ for the cylinder, we see that $X_\pm = \rho x_\pm$ where the $x_\pm$ are the elementary shift operators ($x_\pm |n\rangle = |n\pm 1\rangle$; $X_3|n\rangle = 2\kappa n|n\rangle$ -- see the discussion around (\ref{shift})). Of course, although the action of $X_\pm$ on the fuzzy sphere also raises and lowers the state acted on, because of the relation $[X_+,X_-] = \kappa X_3$, this action is weighted by the Clebsch-Gordon coefficients ($X_\pm |n\rangle = c^n_\pm|n\pm 1\rangle; X_3|n\rangle = \kappa n|n\rangle$). If a solution is to interpolate between these two cases, it would seem that the most obvious form it would take would be the following:

\begin{equation}
\label{ans}
X_+(t) = D(t)x_+ ~,~ X_3(t) = \Delta(t)\tau 
\end{equation}   

Where $x_+$ is the (unitary) elementary shift operator introduced above ($x_+x_- = x_-x_+ = 1$), and $\tau$ is the operator representing the non-compact coordinate on the fuzzy cylinder: $[\tau,x_\pm] = \pm2\kappa x_\pm,~[x_+,x_-] = 0$. That is, we try to deform our cylinder solution into the spherical solution. The operators $D(t)$ and $\Delta(t)$ (and all their time derivatives) are {\it diagonal} with real elements, as $X_3$ only needs its spectrum to be rescaled in making the transition from the the cylinder to the sphere. Similarly, the action of $x_\pm$ also only has to be rescaled so that instead of effecting the shift $|n\rangle \to |n\pm 1\rangle$, the action of this operator becomes that of the raising or lowering operators of su(2). Furthermore, recalling (\ref{lc1})-(\ref{lc2}) we see that it suffices to study the equation of motion for $X_+$ as the equation for $X_-$ is the adjoint of the former (see the discussion following (\ref{gc})). Furthermore lets say that at any given time, we force our ansatz to satisfy the following set of relations:

\begin{equation}
\label{deform}
[X_+(t),X_-(t)] = \lambda(t)X_3(t)~,~[X_3(t),X_\pm(t)] =\pm\theta(t)X_\pm(t)
\end{equation}

Which is to say, our solution will be such that it undergoes a time dependent deformation which takes it from the cylinder to the sphere. With this last qualification, we can straightforwardly evaluate the equations of motion (\ref{lc1})-(\ref{lc3}), resulting in the following:

\begin{equation}
\label{defeom1}
\ddot{X}_+ = (2\kappa - \theta - \lambda/2)\theta X_+  
\end{equation}
\begin{equation}
\label{defeom2}
\ddot{X}_3 = (\kappa - \theta)\lambda X_3  
\end{equation}

Where we suppress all time arguments from here on. One should take care in what follows in realising that $X_-(t) = X_+^\dag(t) = x_-D(t)$, since $D(t)$ is a diagonal operator with only real elements. The form of our ansatz (\ref{ans}) trivially satisfies the Gauss law constraint (\ref{Gauss}) and (\ref{gc}):

\begin{equation}
\label{defgc}
[\dot{\Delta}\tau,\Delta\tau] + \frac{1}{2}\Bigl( [\dot{D}x_+,x_-D] + [x_-\dot{D},Dx_+]\Bigr) = 0
\end{equation}

Where the first term in the above vanishes as both $\Delta$ and $\tau$ are diagonal operators. The terms in the parentheses sum to zero, this time because both $D$ and $\dot{D}$ are diagonal. Hence the structure of our ansatz already takes the Gauss law constraint into account. We note in passing that the Gauss condition for finite N, corresponds to setting all of the $U(N)$ charges to zero. One sees this from calculating the N\"oether charges of (\ref{act2}) under the $U(N)$ symmetry ($X_i \to UX_iU^\dag$), which are given by:

\begin{eqnarray*}
\label{charges}
c^a &=& Tr \bigl(\dot{X}_i[X_i,T^a]\bigr)\\ &=& Tr \bigl(T^a[\dot{X}_i,X_i]\bigr)\\ &=& 0
\end{eqnarray*}

Where the $T^a$ are the generators of $U(N)$. We used the cyclicity of the trace in arriving at the penultimate step above, thus we cannot make this conclusion in the case of infinite dimensional matrices. However this fact is only of passing interest to the problem at hand, with which we will persevere. Requiring the conditions (\ref{deform}) to be satisfied by our ansatz (\ref{ans}) implies the following:

\begin{eqnarray}
\label{defcon1}
[X_+,X_-] = \lambda X_3 ~&\to&~ D^2 - x_-D^2x_+ = \lambda\Delta\tau\\
\label{defcon2}
[X_3,X_+] = \theta X_+ ~&\to&~ D\Delta x_+2\kappa + D(\Delta - x_-\Delta x_+)(\tau - 2\kappa)x_- = \theta Dx_+
\end{eqnarray}
 
Where we repeatedly use the fact that the commutators of diagonal operators vanish, and the relations $[\tau,x_\pm] = \pm 2\kappa x_\pm$ in evaluating the above. The specific form for (\ref{defcon2}) is arrived at by trying to get the the operators $D$ and $x_+$ to the left and right hand sides of each term respectively. We can satisfy the above conditions if the following equations hold true: 

\begin{eqnarray}
\label {dc1} D^2 - x_-D^2x_+ - \lambda\Delta\tau &=& 0\\ \label{dc2} 2\kappa\Delta + (\Delta - x_+\Delta x_-)(\tau - 2\kappa) &=& \theta
\end{eqnarray}          
 
Now realising that the spectrum of $\tau$ is of the form $2\kappa n$, and the nature of the $x_\pm$ as shift operators, we can write down (\ref{dc2}) for each (diagonal) matrix element:

\begin{equation}
\label{wy}
2\kappa\bigl[ \Delta_n + (n-1)(\Delta_n - \Delta_{n-1})\bigr] = \theta  
\end{equation}

Further realising that $X_3 = \Delta\tau$ only needs to undergo a rescaling in going from the sphere to the cylinder ($X_3^{cyl}|n\rangle = 2n\kappa|n\rangle$,  $X_3^{sph}|n\rangle = n\kappa|n\rangle$) we see that more than just being diagonal, picking $\Delta$ to be a time dependent multiple of the identity $\Delta = \omega I$ will do the job. With this in mind, the above condition simplifies greatly and implies:

\begin{equation}
\label{x3}
\omega (t) = \theta/2\kappa ~;~ \Delta = \omega I 
\end{equation} 

Armed with this information, we can now evaluate the constraint (\ref{dc1}), which now implies:

\begin{equation}
\label{x4}
D^2_n - D^2_{n+1} = n\lambda\theta
\end{equation}

Now we know that the job of $D(t)$ is to rescale $x_+$ such that $X_+(t_0) = \rho x_+$ and $X_+(t_f) = J_+$, i.e. to transform the cylinder $X_+$ to the spherical $X_+$, which is the usual raising operator of angular momentum, over a given time interval. Let us take $\rho$ to be equal to the radius of the sphere: $\rho = \kappa\sqrt{N(N+1)}$. Recall that the angular momentum raising operator $J_+ = X_1 + iX_2$ (where $[X_i,X_j] = i\kappa\epsilon_{ijk}X_k$), acts on states as $J_+|n\rangle = \kappa\sqrt{N(N+1) - n(n+1)}|n+1\rangle$. We emphasize that we should suspend our disbelief at the factor $N$ which has started to appear for the moment, as this is a warm up to the more realistic situation we shall consider shortly. We thus suppose a reasonable ansatz for $D(t)$ would be:

\begin{equation}
\label{D}
D_n(t) = \kappa\sqrt{N(N+1) - g(t)n(n-1)}
\end{equation} 

One can convince themselves of the factor $n(n-1)$ (as opposed to $n(n+1)$ which appears in the Clebsch-Gordon coefficients) is appropriate in the ansatz for $D(t)$, by realising that this is the appropriate factor to left multiply $x_+$ in order to scale it to the raising operator. This is easiest seen by writing out this multiplication in a matrix representation. However, one can also see this from the following-- if:

\begin{eqnarray*}
\langle n+1|J_+|n\rangle = \kappa\sqrt{N(N+1) - n(n+1)} ~&,&~ \langle n+1|Dx_+|n\rangle  = D_{n+1}
\end{eqnarray*}

Then at the time at which the transition is complete, $D(t_f)x_+ = J_+$, so that

\begin{eqnarray*}
D_{n+1}(t_f) &=& \kappa\sqrt{N(N+1)- n(n+1)}
\end{eqnarray*}

so that $D_n(t_f) = \kappa\sqrt{N(N+1) - n(n-1)}$, hence our anstatz (\ref{D}). Clearly if $g(t_0) = 0$ and $g(t_f) = 1$, then we satisfy the appropriate initial and final conditions:

\begin{equation}
\label{icfc}
X_{+n}(t_0) = \kappa\sqrt{N(N+1)}x_{+n} ~,~ X_{+n}(t_f) = \kappa\sqrt{N(N+1)-n(n+1)}x_{+n} = J_{+n}
\end{equation}

Hence, (\ref{x4}) implies that:

\begin{equation}
\label{hj}
\frac{\lambda\theta}{2\kappa^2} = g(t)
\end{equation} 

We note that this is a highly non-trivial constraint to consistently satisfy, as in general any one parameter ansatz for a matrix equation has to simultaneously solve an infinite number of equations (for each matrix element), which we fortunately manage in our case by a judicious choice for our ansatz. We observe that (\ref{hj}) is consitent with our initial and final conditions: when $\lambda=0$, $\theta=2\kappa$ which corresponds to the cylinder, $g=0$. When $\lambda=2\kappa$, $\theta=\kappa$ which corresponds to the sphere, $g=1$ (c.f. (\ref{c})(\ref{s}) and (\ref{deform})). Having satisfied both constraints in (\ref{deform}), we are ready to solve for the equations of motion. Before we do this, we summarize our findings:

\begin{eqnarray}
\label{tv1}
\Delta(t) &=& \theta(t)/2\kappa ~;~ X_3(t) = \frac{\theta(t)}{2\kappa}\tau\\
\label{tv2}
D_n(t) &=& \kappa\sqrt{N(N+1) - g(t)n(n-1)}~;~ g(t) = \frac{\lambda(t)\theta(t)}{2\kappa^2}
\end{eqnarray}
 
Thus our equations of motion (\ref{defeom1}) and (\ref{defeom2}) rather straighforwardly imply the following equations for our parameters:

\begin{eqnarray}
\label{teom}
\ddot{\theta} &=& 2(\kappa - \theta)\kappa^2g\\ 
\label{geom}
\ddot{D}_n &=& [(2\kappa - \theta) - \kappa^2g]D_n
\end{eqnarray}

We see that in general, (\ref{geom}) could be inconsistent-- writing out explicitly what this equation implies:

\begin{equation}
\label{Ddot}
\frac{\ddot{D}_n}{D_n} = -\frac{\ddot{g}n(n-1)\kappa^2}{2D^2_n} -\frac{\dot{g}^2n^2(n-1)^2\kappa^4}{4D^4_n}~=~ [(2\kappa-\theta)\theta - \kappa^2g]
\end{equation} 

We see that the left hand side depends on $n$, whereas the right hand side does not. Consider the factor that appears in either term on the left:

\begin{equation}
\label{ggb}
\frac{n(n-1)\kappa^2}{D^2_n} = \frac{1}{\bigl[\frac{N(N+1)}{n(n-1)} - g\bigr]} \to 0 ~as~ N \to \infty
\end{equation}

Because of this, we might think that since this model is only valid in the large N limit, both sides of (\ref{Ddot}) have to vanish (which ensures that the $n$ dependence of the equation is inconsequential). However this is an incorrect conclusion. N would truly have to be infinite if we are to even admit the fuzzy cylinder as a solution, and this calls into question the very consistency of what we are doing, which we remind the reader again is simply warming up for a more realistic treatment. We see that if this equation is to be satisfied for all finite $n$, then a safe way to do it would be to demand that $g$ evolve adiabatically: $\dot{g}^2$ , $\ddot{g} \ll 1$. That is, if we assume the adiabatic behaviour of our ansatz, the left, and consequently the right hand side of (\ref{Ddot}) would have to be vanishingly small:

\begin{equation}
\label{gga}
(2\kappa - \theta)\theta = \kappa^2g + \epsilon
\end{equation}   

Which we can safely take to imply the following for (\ref{teom}):

\begin{equation}
\label{tfeom}
\ddot{\theta} = 2(\kappa - \theta)(2\kappa - \theta)\theta
\end{equation} 

Certainly we would not have to be too fussy about the adiabatic condition if we are only concerned with the evolution of matrix elements at finite $n$, if we took $N$ to be very large (but still finite). This is because the factors (\ref{ggb}) multiplying the time derivatives of $g$ are highly suppressed. Expressing $\theta$ in terms of dimensionless variables ($\tilde{\theta} = \theta/\kappa$), the equations above become: 

\begin{eqnarray}
\label{tt1} \ddot{\tilde{\theta}} &=& 2\kappa^2(1-\tilde{\theta})(2-\tilde{\theta})\tilde{\theta}\\ \label{tt2}g &=& \tilde{\theta}(2-\tilde{\theta}) 
\end{eqnarray}

From which we see that if (\ref{tt1}) and (\ref{tt2}) are to be consistent with each other, and the adiabatic condition, then neccesarily $\kappa^2 \ll 1$. That is, through the relation (\ref{tt2}), if $g$ is to vary slowly, then so must $\tilde{\theta}$. This is only possible for weak coupling to the RR background, which is equivalent to the commutative limit of the model. We will have more to say on this further on. The equation of motion for $\tilde{\theta}$ (\ref{teom}) is equivalent to motion in the potential:

\begin{equation}
\label{pot}
V(\tilde{\theta}) = 2\kappa^2(\tilde{\theta}^3 - \tilde{\theta}^2 - \frac{\tilde{\theta}^4}{4})
\end{equation}

\begin{figure}
\epsfig{figure=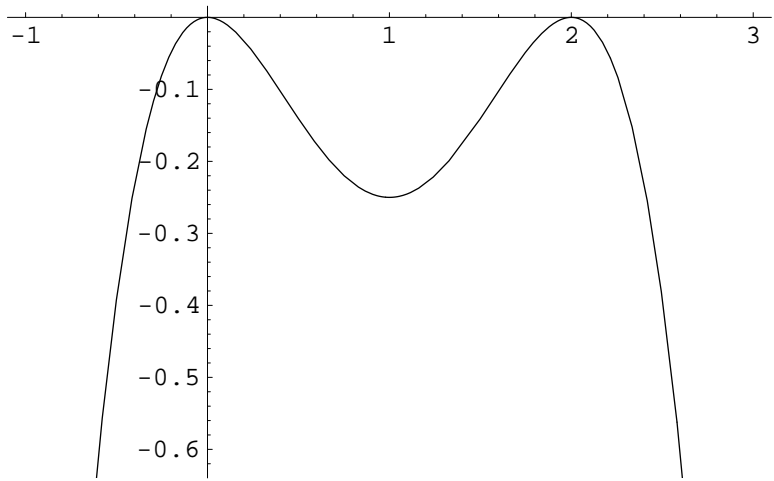}
\caption{$V/2\kappa^2$ plotted as a function of $\tilde{\theta}$}
\end{figure}

Which we plot in Fig. 1. We note that we are only interested in the behaviour of the model for $\tilde{\theta}$ ranging between 0 and 2, the regions exterior to this domain having been included to illustrate the rather generic `wine bottle' shape of this potential. If this model is to be taken seriously, we can immediately discern several salient features of the solution from this potential. Firstly, since this ansatz is only consistent in the limit $\kappa^2 \ll 1$, the energies of the spherical and cylindrical solutions (\ref{compa}) tend towards degeneracy with that of the commuting solution (a gas of D0 branes), and hence should freely transmute into one another. From the potential, we see that we have uncovered the marginal instability in this limit of the commuting and cylindrical solutions (which correspond to $\tilde{\theta} = 2$ and $0$ respectively)-- the instability is parametrized by the curvature at the maxima, which depends on $\kappa^2$, and is hence very slight. We note in passing that this treatment says nothing of the stability of the cylindrical solutions for finite values of $\kappa^2$. We also see that the spherical solution (which corresponds to $\tilde{\theta} = 1$) is still a marginally bound state. We can interpret the dynamics resulting from the anstaz in two ways-- either as an instanton transition between a gas of D0 branes and the fuzzy cylinder, (which would correspond to motion between $\tilde{\theta} = 0$ and $\tilde{\theta} = 2$), or one could only follow the ansatz only up until the fuzzy sphere has been formed. Clearly unless there is some dissipative mechanism involved, the resulting post transition fuzzy sphere geometry cannot be static. However, if at the point $\tilde{\theta} = 1$, we were to drop all the restrictions (\ref{ans})(\ref{deform}) and follow the solutions from then on, it is not inconcievable that a spherical solution might persist, except now as some undulating deformation of the sphere. Whatever the final outcome, were this admitedly unrealistic example to be taken seriously, it seems that we have modelled topological transitions using matrix dynamics in the large N limit. Of course any topology change would be occuring at infinity (spatially, as well as in terms of matrix elements) in this model, where it distinctly inconsistent to draw any firm conclusions. However we find that having acquainted ourselves with the nature of the problem, we only need to make a few modifications to this set up to model a much more realistic situation. We find that in the case where we study the decay of the cylinder into an infinite collection of spheres, we come across many of the features we just uncovered in this toy example but in more grounded circumstances, which is where we turn our attention towards now.

\section{The Decay of the Fuzzy Cylinder: A Classical Transition?}    

We note that although the conditions (\ref{deform}) facilitated the subsequent calculations, they may overly restirct our problem. They were proposed so that at each point along the trajectory of the solution, we had a firm interpretation of the intermediate state as a deformation of the cylinder. Since we are proposing a solution which transmutes bewteen the cylinder and an infinite collection of spheres (see fig.2), it seems reasonable to suppose that one may not have such an interpretation available in this case. It turns out that if we were to drop this condition at the outset, the equations of motion would imply them for matrix elements which are not involved (in a sense we will make precise later) with the topology change, whereas this will not be the case for the matrix elements which are directly involved in this transition. We find that dropping the conditions (\ref{deform}) does not reduce the amount of work we have to do in any sense, as the difficulty in constructing a solution is shifted from trying to make our ansatz consistent with this condition, to evaluating the equations of motion, which these conditions simplified. Before we can proceed, we take note of an important aspect of the fuzzy sphere solutions that we've glossed over till now. Consider the static equations of motion:

\begin{figure}
\epsfig{figure=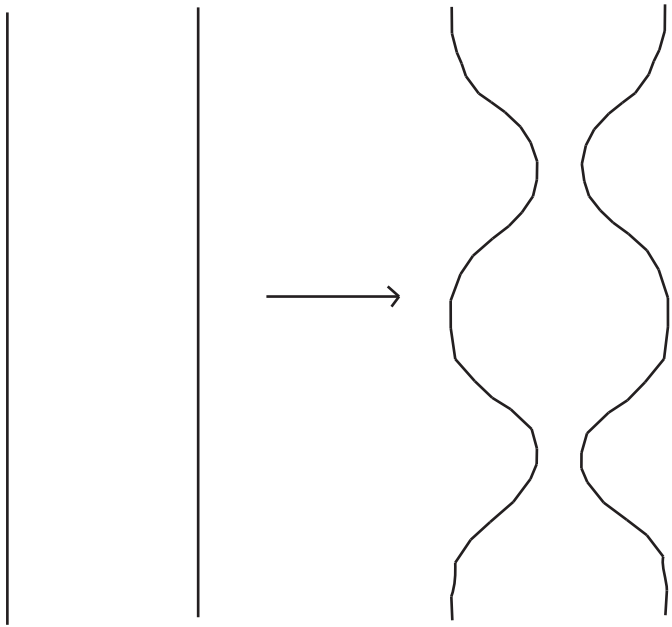}
\caption{Schematic representation of the transition at intermediate times}
\end{figure}

\begin{equation}
[X_j,[X_i,X_j] - i\kappa\epsilon_{ijk}X_k] = 0
\end{equation}

We note that the fuzzy sphere solution given by $[X_i,X_j] = i\kappa\epsilon_{ijk}X_k$ admits the generalization \cite{k4}:

\begin{equation}
\label{marg}
X_i = \kappa J_i + Y_i
\end{equation}

Where the $Y_i$ are any operators that commute with the generators of $su(2)$ given by the $J_i$ and themselves:

\begin{equation}
\label{m2}
[Y_i,J_k] = 0~,~ [Y_i,Y_j] = 0
\end{equation}

In particular, if we consider the reducible solution given by 
$X_i = \oplus_{r} \kappa J_i^{(r)}$, where the sum is over irreducible representations, then an obvious operator that satisfies (\ref{m2}) would be given by:

\begin{equation}
\label{y}
Y_i = \oplus_{r} c_i^{(r)} I^{(r)}
\end{equation}

Where the $c_i^{(r)}$ are arbitrary coefficients weighting the $I^{(r)}$, which are the identity operators on the subspaces defined by the reducible representation labelled by $r$. These so called marginal deformations do not change the energy of the solutions, and correspond to center of mass dispacements of the individual fuzzy spheres that make up the solution. One can then easily construct a static situation where these center of mass displacements are such that we have an infinite tower of spin j fuzzy spheres along the $X_3$ direction. This is the solution that we will attempt to connect to the cylinder. To facilitate the process of trying to construct our ansatz, let us consider what the two solutions we are trying to dynamically relate look like in a matrix representation. When our solution describes the cylinder (at the initial time), the operator $X_+$ takes on the form:

\begin{equation}
\label{cylmat}
X_+ = \rho{\begin{pmatrix} \ddots&\vdots&\vdots&\ddots\\
\ldots&{\begin{vmatrix} 0&1&0&\ldots&0&0\\0&0&1&\ldots&0&0\\ \vdots&\vdots&\vdots&\ddots&\vdots&\vdots \\0&0&0&\ldots&1&0\\0&0&0&\ldots&0&1\\0&0&0&\ldots&0&0
\end{vmatrix}}&
{\begin{vmatrix} 0&0&0&\ldots&0&0\\0&0&0&\ldots&0&0\\ \vdots&\vdots&\vdots&\ddots&\vdots&\vdots \\0&0&0&\ldots&0&0\\0&0&0&\ldots&0&0\\1&0&0&\ldots&0&0
\end{vmatrix}}&\ldots\\
\ldots&{\begin{vmatrix} 0&0&0&\ldots&0&0\\0&0&0&\ldots&0&0\\ \vdots&\vdots&\vdots&\ddots&\vdots&\vdots \\0&0&0&\ldots&0&0\\0&0&0&\ldots&0&0\\0&0&0&\ldots&0&0
\end{vmatrix}}&
{\begin{vmatrix} 0&1&0&\ldots&0&0\\0&0&1&\ldots&0&0\\ \vdots&\vdots&\vdots&\ddots&\vdots&\vdots \\0&0&0&\ldots&1&0\\0&0&0&\ldots&0&1\\0&0&0&\ldots&0&0
\end{vmatrix}}&\ldots\\
\ddots&\vdots&\vdots&\ddots
\end{pmatrix}}
\end{equation}   

Where we have partitioned the matrix into $N$ by $N (= 2j+1)$ blocks. The matrix representation of $X_+$, when it describes an infinite tower (along the $X_3$ axis) of spin j spheres is given schematically as:

\begin{equation}
\label{sphmat}
X_+ = \kappa{\begin{pmatrix} \ddots&\vdots&\vdots&\ddots\\
\ldots&{\begin{vmatrix} 0&c^{j-1}_+&0&\ldots&0&0\\0&0&c^{j-2}_+&\ldots&0&0\\ \vdots&\vdots&\vdots&\ddots&\vdots&\vdots \\0&0&0&\ldots&c^{-j+1}_+&0\\0&0&0&\ldots&0&c^{-j}_+\\0&0&0&\ldots&0&0
\end{vmatrix}}&
{\begin{vmatrix} 0&~~0~~&~~0~~&\ldots&~~0~~&~~0~~\\0&0&0&\ldots&0&0\\ \vdots&\vdots&\vdots&\ddots&\vdots&\vdots \\0&0&0&\ldots&0&0\\0&0&0&\ldots&0&0\\0&0&0&\ldots&0&0
\end{vmatrix}}&\ldots\\
\ldots&{\begin{vmatrix} 0&~~0~~&~~0~~&\ldots&~~0~~&~~0~~\\0&0&0&\ldots&0&0\\ \vdots&\vdots&\vdots&\ddots&\vdots&\vdots \\0&0&0&\ldots&0&0\\0&0&0&\ldots&0&0\\0&0&0&\ldots&0&0
\end{vmatrix}}&
{\begin{vmatrix} 0&c^{j-1}_+&0&\ldots&0&0\\0&0&c^{j-2}_+&\ldots&0&0\\ \vdots&\vdots&\vdots&\ddots&\vdots&\vdots \\0&0&0&\ldots&c^{-j+1}_+&0\\0&0&0&\ldots&0&c^{-j}_+\\0&0&0&\ldots&0&0
\end{vmatrix}}&\ldots\\
\ddots&\vdots&\vdots&\ddots
\end{pmatrix}}
\end{equation}   

Where $c^{n}_+$ are the Clebsch-Gordon coefficients given by $c^{n}_+ = \sqrt{j(j+1) - n(n+1)}$. More succinctly, we can say that the when $X_+$ is describing an infinite tower of spin j spheres, it is a block diagonal series of the spin j raising operator $J_+$. We take the only nonzero deformation $Y_i$ to be $Y_3$, which will have the general form (\ref{y}). Namely, the final collection of fuzzy spheres will all be displaced along the z direction. Since $X_3$ will be a diagonal operator in the initial and final state, we can assume that the full time dependence of the operator will be given as before:

\begin{equation}
\label{del}
X_3(t) = \Delta(t)\tau
\end{equation}  

Where $\tau$ is the non-compact coordinate of the cylinder, with $\Delta$ also taken to be diagonal. If one were to consider the $j\times j$ matrix which transforms the diagonal blocks in (\ref{cylmat}) to those in (\ref{sphmat}):

\begin{equation}
\label{dmat}
\rho{\begin{pmatrix} 0&1&0&\ldots&0&0\\0&0&1&\ldots&0&0\\ \vdots&\vdots&\vdots&\ddots&\vdots&\vdots \\0&0&0&\ldots&1&0\\0&0&0&\ldots&0&1\\0&0&0&\ldots&0&0
\end{pmatrix}} ~\to~ \kappa{\begin{pmatrix} 0&c^{j-1}_+&0&\ldots&0&0\\0&0&c^{j-2}_+&\ldots&0&0\\ \vdots&\vdots&\vdots&\ddots&\vdots&\vdots \\0&0&0&\ldots&c^{-j+1}_+&0\\0&0&0&\ldots&0&c^{-j}_+\\0&0&0&\ldots&0&0
\end{pmatrix}}
\end{equation}

We see that an ideal candidate would be the matrix $d(t)$, similar to the one defined previously but now, a finite dimensional matrix (recall that if we set the radius of the cylinder to that of any of the individual spin j spheres we are trying to evolve to, then $\rho = \kappa\sqrt{j(j+1)}$):

\begin{equation}
\label{dmatt}
d_m(t) := \kappa\sqrt{j(j+1) - g(t)m(m-1)}~;~ -j \leq m \leq j
\end{equation}
 
Notice that $d_{-j} = \kappa\sqrt{j(j+1)(1 - g(t))}$ seems to be a redundant element of our definition, as when we multiply the initial block matrix in (\ref{dmat}) by $d(t)$, it's $-j^{th}$ element does not multiply any non-zero elements and hence drops out. However it serves the very important function of scaling the off-diagonal block elements in (\ref{cylmat}) to zero at the end of the evolution (when $g = 1$)-- the non zero elements being in the upper off-diagonal blocks. That is, if we were to define the operator $D(t)$ as the direct sum of an infinite number of these $(2j+1)\times(2j+1)$ block matrices, we see that the time dependent ansatz:

\begin{equation}
\label{dans}
X_+(t) = D(t)x_+ ~;~ D(t) = diag(...d(t),d(t),d(t)...)
\end{equation}

can scale (\ref{cylmat}) to (\ref{sphmat}), where $x_+$ as before is the elementary shift operator (see discussion surrounding (\ref{ans})). The equations of motion (\ref{lc1}) and (\ref{lc3}) now become:

\begin{equation}
\label{feom1}
\ddot{\Delta}\tau = [(\kappa - \Omega)D^2 - x_-(\kappa - \Omega)D^2x_+]
\end{equation}

\begin{equation}
\label{feom2}
\ddot{D} = D\bigl[ \frac{1}{2}(x_+D^2x_- + x_-D^2x_+ - 2D^2) + (2\kappa - \Omega)\Omega]
\end{equation}

\begin{equation}
\label{feom3}
\Omega := [2\kappa\Delta + (\Delta - x_+\Delta x_-)(\tau - 2\kappa)]
\end{equation}

Obtaining these is a straightforward excercise, where we only use the cylinder relations $[\tau,x_\pm] = \pm2\kappa x_\pm$, and the fact that the operators $\Delta$ and $D$ are diagonal (recall from the discussion around   
(\ref{defgc}) that the Gauss law constraint (\ref{gc}) is trivially satisfied in this case). Evaulating the above for specific forms for these scaling operators in a consistent way is then the remaining challenge. Before we do this, we outline a few preliminaries that will help us further on. Since we are aiming to transmute the cylinder solution into an infinite collection of spin j fuzzy spheres, each displaced along the $x_3$ axis, it should be clear that the final form of the solution should be something like:

\begin{equation}
\label{x3f}
X_3(t_f) = \oplus_r (\kappa J^r_3 + c^r_3 I^r) 
\end{equation} 
    
Where the $I^r$ are the identity operators on the $2j+1$ dimensional subspaces defined by our decomposition, and the $c^r_3$ are some as yet undetermined displacements. Consider now the following counting scheme for basis vectors in our Hilbert space-- let any value for the index $n$ be given by

\begin{equation}
\label{ord}
n = (2j+1)w + m ~~;~~ w ~\epsilon~ \mathbb Z~, -j \leq m \leq j
\end{equation}

That is, $w$ labels the block we are studying and $m$ represents the index within that block. Consider now the operator $\Omega$ (\ref{feom3}), if we define $\Delta$ as (c.f. our ansatz in the previous section):

\begin{equation} 
\label{dela}
\Delta_n(t) =  \frac{\theta(t)}{2\kappa} + \frac{\omega(t)w}{2\kappa n}
\end{equation}

where $\theta$ and $\omega$ are as yet unspecified functions of time, and $w$ implicitly depends on $n$ from (\ref{ord}). It is straightforward to see effect of this definition on $X_3(t)$:

\begin{equation}
\label{x3t}
X_{3n}(t) = \Delta_n(t)\tau_n = \frac{\theta}{2\kappa}\tau_n + \omega w  
\end{equation} 

That is the effect of $\Delta$ is to rescale the cylinder operator, in addition to imparting a block dependent translation (this translation is independent of the index $m$ within the block-- see (\ref{ord})). This is precisely what we expect of an ansatz that will evolve to (\ref{x3f}). Thus with this definition, we can evaluate $\Omega$ as:

\begin{eqnarray*}
\label{om}
\Omega_n = \Omega_{(2j+1)w+m} &=& \theta ~;~ m \neq -j\\ &=& \theta + \omega ~;~ m= -j
\end{eqnarray*}

Where the fact that $w$ changes when we move from one block to another accounts for the difference when $n$ is such that $m = -j$ (i.e. we are considering the index corresponding to the lowest weight state within that block). Recalling the definition of $D$ from (\ref{dmatt}) and (\ref{dans}), it is a straight forward matter to see that for $n$ such that $m \neq \pm j$, our ansatz satisfies (\ref{feom1}) if:

\begin{eqnarray}
\label{asg1}
\ddot{\theta} &=& 2\kappa^2g(\kappa - \theta)\\
\label{asg2}
\ddot{\omega} &=& -(2j+1)\ddot{\theta} 
\end{eqnarray}   

These conditions also imply that (\ref{feom1}) is satisfied When $n$ is such that $m = -j$. However in the case when $m = j$, (\ref{feom1}) explicitly reads:

\begin{equation}
\label{asg3}
[(2j+1)w + j]\ddot{\theta} + \ddot{\omega}w = (\kappa - \theta)D^2_n - (\kappa - \theta - \omega)D^2_{n+1} 
\end{equation} 

Which implies (recalling that the index $w$ is one higher for the state $n+1$ when $n$ is such that $m = j$, that is the state $n+1$ resides in the next highest block representation):

\begin{equation}
\label{asg4}
\omega\kappa^2j(j+1)(1-g) + 2j(\kappa - \theta)\kappa^2g = (2j+1)w\ddot{\theta} + w\ddot{\omega} + j\ddot{\theta}
\end{equation}

The conditions (\ref{asg1}) and (\ref{asg2}) imply that several cancellations occur, leaving us with the condition:

\begin{equation}
\label{dcond}
\omega\kappa^2j(j+1)(1-g) = 0
\end{equation} 

Before we can interpret this, we use (\ref{asg2}) to solve for $\omega$ consistent with the initial conditions $\theta(t_0) = 2\kappa$, $\omega(t_0) = 0$ (which come from requiring $\Delta(t_0) = I~ (\ref{dela}))$:

\begin{eqnarray*}
\label{om2}
\omega &=& -(2j+1)\theta + 2\kappa(2j+1)\\
&=& (2j+1)(2\kappa - \theta)
\end{eqnarray*}

From which we conclude that the above condition becomes:

\begin{equation}
\label{dcondi}
\kappa^2(2J+1)j(j+1)(2\kappa - \theta)(1-g) = 0
\end{equation}

Which is not going to be satisfied at intermediate times in our evolution. However we will see further on that a limit which we will be compelled to take will ensure that this quantity is negligible. We are still left with one more matrix equations of motion (\ref{feom2}) to satisfy. Evaluating this for $n$ such that $m \neq \pm j$, we get:

\begin{equation}
\label{sas1}
\ddot{D}_n = D_n[(2\kappa - \theta)\theta - \kappa^2g]
\end{equation}

We recognize this immediately as (\ref{geom}) from our previous section. We can argue just as we did then, namely that since this equation implies the following:

\begin{equation}
\label{Ddott}
\frac{\ddot{D}_n}{D_n} = -\frac{\ddot{g}m(m-1)\kappa^2}{2D^2_n} -\frac{\dot{g}^2m^2(m-1)^2\kappa^4}{4D^4_n}~=~ [(2\kappa-\theta)\theta - \kappa^2g]~~; n = (2j+1)w + m~,~m \neq \pm j
\end{equation} 

That in order for this equation to be saisfied for all $m \neq \pm j$, our evolution should be adiabatic ($\ddot{g} \ll 1$, $\dot{g}^2 \ll 1$), and the right hand side should be vanishingly small \footnote{Note that we need not consider this equation for $m=0,1$ as the time derivatives of $D_n$ vanish for these elements.}. That is, just as in the previous section:

\begin{equation}
\label{id}
\kappa^2g - (2\kappa - \theta)\theta = \epsilon
\end{equation}  

which we can safely take to imply for the equation of motion (\ref{asg1}) the following (defining $\tilde{\theta} = \theta/2\kappa$):

\begin{eqnarray}
\label{ttt1} \ddot{\tilde{\theta}} &=& 2\kappa^2(1-\tilde{\theta})(2-\tilde{\theta})\tilde{\theta}\\ \label{ttt2}g &=& \tilde{\theta}(2-\tilde{\theta}) 
\end{eqnarray}

Which are exactly the same equations as (\ref{tt1})(\ref{tt2}), which implied that our entire ansatz could be boiled down to the behaviour of a single parameter which appeared as if it was moving in the potential (\ref{pot}) indicated in fig. 1. Hence it appears that in order for our ansatz to be self-consistent, the adiabatic condition translates into the fact that our RR field strength $\kappa$ must be very small. However we run into an inconsistency with this demand. Examining the factors weighting the time derivatives of $g$ in (\ref{Ddott}):

\begin{equation}
\label{milla}
\frac{m(m-1)\kappa^2}{D^2_m} = \frac{1}{\bigl[\frac{j(j+1)}{m(m-1)} - g\bigr]}
\end{equation}

We see that we cannot safely assume that these vanish uniformly. Recall that since in the previous section, $N$ took the place of $j$ which in the (unrealistic) infinite $N$ limit (provided we looked at matrix elements indexed by a finite value of $n$), this factor would vanish. Since we are attempting to study the transition to finite dimensional spheres, we see that in general this factor will not be very small for a great many matrix elements. This would not be a problem if the time derivatives of $g$ were smaller than any other scale of interest in the problem, since it is the product of the time derivates with these factors that we'd like to be vanishingly small. However in this problem, since the right hand side of (\ref{Ddott}) is taken to vanish almost completely (and independently) through the identification (\ref{ttt2}), the adiabatic condition implies that $\ddot{g}$, $\dot{g}^2 \sim\kappa^2$ (\ref{ttt1})(\ref{ttt2}). Hence to ignore this would be inconsistent since we're clearly considering terms of this order in (\ref{ttt1}). However this situation becomes even worse when we consider the behaviour of the factor (\ref{milla}) when $m \to \pm j$:

\begin{equation}
\label{milla2}
\frac{m(m-1)\kappa^2}{D^2_m} = \frac{1}{\bigl[\frac{j(j+1)}{m(m-1)} - g\bigr]} ~\to~ \frac{1}{(1 - g)}
\end{equation}

which clearly diverges as $g \to 1$ \footnote{We wish to point out that were we to consider the equations of motion arising from (\ref{feom2}) for the case when $m = j$, we arrive at the same conditions as above (\ref{ttt1})(\ref{ttt2}). In the case when $m = -j$, we find that the above conditions will only solve the equations of motion up to a term of order $\kappa^2$-- which we cannot take to vanish. Hence at $m = -j$ in addition to the divergence of the factor (\ref{milla2}), we have an additional contribution to the faliure of this ansatz.}. This corresponds to points in our trajectory becoming inconsistent with the equations of motion as we near the spherical solution. That is even if we could somehow enforce the adiabatic condition consistently, the approximation breaks down violently at matrix elements that are near the poles of the fuzzy spheres we are trying to evolve to. That is, it appears as if there is no way to ensure a consistent solution to (\ref{feom2}) with this ansatz. Recall that what saved us from running into this problem in the case of a (fictitious) infinite dimensional representation of the sphere was the fact that the factor (\ref{ggb}), for finite $n$ was extremely small in the large $N$ limit. However such considerations completely ignore the matrix elements which are somehow involved in the topology changing process. We see that in the more realistic example of studying the decay of the fuzzy cylinder into an infinite collection of fuzzy spheres, our ansatz simply does not work as a solution to the classical equations of motion up to order $\kappa^2$. However upon closer examination, we find that this attempt yields to several interesting resolutions, which we address now.

\section{Topology Change: Contact Interactions and Quantum Transitions}

In order to have a clearer train of thought in this section, we briefly summarize the findings of the last section. We started with the observation that the ansatz given by:

\begin{eqnarray*}
X_+(t) = D(t)x_+ &~;~& D(t) = diag(...d(t),d(t),d(t)...)\\ &~& d_m(t) := \kappa\sqrt{j(j+1) - g(t)m(m-1)}\\ X_3(t) = \Delta(t)\tau &~;~&
\Delta_n(t) =  \frac{\theta(t)}{2\kappa} + \frac{\omega(t)w}{2\kappa n}\\ &~& n := (2j+1)w + m
\end{eqnarray*}

satisfied the equations of motion (\ref{feom1})-(\ref{feom3}) for almost all matrix elements if the we made the identifications (\ref{asg1})(\ref{asg2})(\ref{ttt2}):

\begin{eqnarray*}
\ddot{\theta} &=& 2\kappa^2g(\kappa - \theta)\\
\ddot{\omega} &=& -(2j+1)\ddot{\theta}\\ g &=& \tilde{\theta}(2-\tilde{\theta})\\ 1 &\gg& \ddot{g} ~,~ \dot{g}^2 
\end{eqnarray*}   

This corresponded to the following equation of motion for $\tilde{\theta} = \theta/2\kappa$:

\begin{eqnarray*}
\ddot{\tilde{\theta}} &=& 2\kappa^2(1-\tilde{\theta})(2-\tilde{\theta})\tilde{\theta}
\end{eqnarray*}  

Which implied that in order for our ansatz to be consistent, $\kappa^2 \ll 1$. That is, our entire ansatz could be boiled down to the motion of a single parameter $\tilde{\theta}$ in the potential (\ref{pot}) given in figure 1. However we found that this ansatz violated the equation of motion for $X_3$ (\ref{feom1}) for matrix elements indexed by $n$ such that $m = -j$, up to a term of order $\kappa^3$ (\ref{dcondi}). That is, this ansatz fails to solve the equations of motion for $X_3$ at matrix elements which are the lowest states of a given block representation, up to order $\kappa^3$. This should not concern us too much in light of the condition $\kappa^2 \ll 1$. However we found that according to (\ref{Ddott}), in attempting to satisfy the equation of motion for $X_+$ (\ref{feom2}), we could only ensure consistency with our ansatz for matrix elements such that the factor (\ref{milla}) given by:

\begin{eqnarray*}
\frac{m(m-1)\kappa^2}{D^2_m} &=& \frac{1}{\bigl[\frac{j(j+1)}{m(m-1)} - g\bigr]}
\end{eqnarray*}
   
happen to be vanishingly small. This arises from the realisation that although the right hand side of (\ref{Ddott}) does not depend on $n$, the left hand side does, unless both sides of the equation vanish (or are vanishingly small). We attempted this for the right hand side through the identification $\kappa^2g - \theta(2-\theta) = \epsilon$, where we take $\epsilon$ to be small independent of $\kappa$. As for the left hand side, we require all time derivatives of $g$ to be small, but the best we could manage consistent with the equations of motion for $\theta$(\ref{ttt1})(\ref{ttt2}) was $\dot{g}^2$, $\ddot{g} \sim \kappa^2$, which also has to be taken to be small. However since we are clearly not neglecting terms of order $\kappa^2$, we need the factors multiplying $\dot{g}^2$ and $\ddot{g}$ in (\ref{Ddott}) to vanish. However these factors are not unifromly small, and for matrix elements close to the poles of the block representations we are considering ($m \to \pm j$), this factor behaves as:

\begin{equation}
\label{po}
\frac{m(m-1)\kappa^2}{D^2_m} = \frac{1}{\bigl[\frac{j(j+1)}{m(m-1)} - g\bigr]} \to \frac{1}{1-g}
\end{equation}
    
Which diverges as we head towards the spherical solution ($g \to 1$)-- even within the adiabatic approximation we cannot satisfy the equations of motion for matrix elements that approach the highest or lowest weight states within a block. Hence in the limit that $\kappa^2 \ll 1$ our ansatz satisfies the matrix equations of motion only for those matrix elements of (\ref{feom2}) which do not represent regions close to the poles of the spin j spheres we are trying to evolve to. However the matrix equation (\ref{feom1}) is satisfied consistently by our anstaz for all matrix elements, the only violations occuring exactly at the poles and are of negligable magnitude in the above limit.\linebreak    

\par

Although our ansatz fails as a solution to the classical equations of motion, we find that the manner in which it fails to be rather interesting. Reconsider (\ref{Ddott}):

\begin{equation}
\label{Ddottt}
\frac{\ddot{D}_n}{D_n} = -\frac{\ddot{g}m(m-1)\kappa^2}{2D^2_n} -\frac{\dot{g}^2m^2(m-1)^2\kappa^4}{4D^4_n}~=~ [(2\kappa-\theta)\theta - \kappa^2g]
\end{equation}

Our ansatz failed to solve this equation of motion only for the matrix elements close to the poles ($m = \pm j$), and for times such that $g \to 1$. Since we must have $\dot{g}^2$, $\ddot{g} \sim \kappa^2$, we see that in this limit, the left hand side of the above diverges as (\ref{milla2}):

\begin{equation}
\label{div}
\frac{\kappa^2}{(1-g)^2}
\end{equation} 

Suppose now we asked the question, what addition to our equations of motion would make our ansatz consistent? We see that in the limit $\kappa^2 \to 0$, with $g$ such that $g(t_0) = 0$, $g(t_f) = 1$, (\ref{div}) can be made to converge to the functional (if the limit $\kappa^2 \to 0$ is taken in a particular way):

\begin{equation}
\label{dirac}
\frac{\kappa^2}{(1-g)^2} \to \lambda \delta(t - t_f)
\end{equation} 

Where $\lambda$ is some (finite) constant. So if instead of the equation of motion (\ref{lc1}) (which gave rise to equation (\ref{feom2})) we had the equation of motion:

\begin{equation}
\label{llc11}
\ddot{X}_+ ~= ~\frac{1}{2}[X_+, [X_+,X_-] ] ~+ ~[X_3, [X_+,X_3] + 2\kappa 
X_+ ] ~+\lambda\delta(t - t_f)MX_+ 
\end{equation} 

where $M$ is a diagonal matrix such that $M_n = 1$ if $n$ is such that $m = \pm j$ and $M_n = 0$ otherwise, we would have in addition precisely this delta function term on the right hand side of (\ref{Ddottt}). Our ansatz would now be a consistent solution to this new equation of motion. One could interpret this as arising from a contact (Yukawa type) interaction added to the action (\ref{act}):

\begin{equation}
\label{interact}
S_{int} = \int dt ~ Tr (X_iY_{ij}X_j)
\end{equation}

With $Y_{3i} = Y_{12} = Y_{21} = 0$, $Y_{11} = Y_{22} = \lambda\delta(t - t_f) M$ giving us our desired modification (since the equation of motion for $X_3$ (\ref{feom1}) is satisfied by our ansatz, we would like $Y_{3i}$ to vanish so as not to alter this equation). Although this might seem to be an ad-hoc imposition, it is not inconceivable that such an interaction might arise from more fundamental string theoretical considerations. Certainly considering the effect of additional background fields assuming a non-zero value in (\ref{act}) might effect this type of term. For instance, were the dilaton background to assume a non-zero expectation value, a term like the one above would be induced, except in this case, one would have:

\begin{equation}
\label{dvev}
Y_{ij} = m^2_{\phi}\delta_{ij}I
\end{equation}

Although the form of the operator $M$ might seem a bit contrived, it is in fact far from contrived. All this extra interaction term serves to do is to provide a source like term for the matrix elements which lie at the poles of the spheres, and as such could be considered as particles which mediate the topology change, or perturbations which cause the topology change. If one were to consider the action of the Yukawa interaction as a series of perturbations, we see that these perturbations are what cause the `rips' to occur.  There is also a suggestive analogy between this and the action of vertex operators on the string worldsheet, but this analogy at present is nothing more than qualitative-- our matrix model has very little to do with conformal field theory in this guise. Hence, the main conclusion we care to draw from this is that topology change (as far as this ansatz is concerned) requires our model to be perturbed somehow, either in an ad-hoc manner or through the mediation of some other degrees of freedom. Nevertheless, we feel that as far as uncovering the existence of topological dynamics in this model as a classical phenomenon, our ansatz seems to have risen to the task.\linebreak

\par

We take note of some points that we've been somewhat cavalier about in the treatment above. Careful consideration of (\ref{Ddottt}) shows that in general, if $j$ is very large, then the factor behaves as (\ref{po}) for many matrix elements near the poles of our block representation. Of course one could always modify the operator $M$ to account for this, but this would make the problem start to look rather contrived. One thus concludes that the spin $j$ spheres we are trying to evolve to cannot be too big. Thus we have a play-off between the limit $\kappa^2 \to 0$ and the fact that $j$ cannot be too large. Whereas the former is trying to make the resultant spheres commutative, the latter is trying to keep the discrete structure manifest. The other fact that we've been somewhat cavalier about is that in order for the considerations going into (\ref{dirac}) to be consistent (and for the violations of (\ref{Ddottt}) to truly occur only close to the poles), we really need $\kappa^2$ to be very small indeed. Since the non-comutativity parameter in the structure relations is always $\kappa$, we will still be dealing with fuzzy geometries, but only in a limit where the parameter tends towards the commutative limit. Note that there is no apparent paradox in our findings-- topology change as we've uncovered it is a distictly non-commutative phenomenon. The main parameter that governs the physical non-commutativity of any fuzzy geometry is the dimension of the matrices being dealt with, as the algebra of $N\times N$ matrices will tend to the algebra of $C^\infty$ functions only in the infinite $N$ limit. Hence we can interpret the limit taken in our ansatz as that corresponding to a marginally commutative cylinder decaying into a series of distinctly non-commutative spheres.\linebreak 

\par     

In addition to the interpretation of our ansatz as describing classical topology change in the situation where our model is acted on by a contact interaction with some external field, one has an interpretation of this solution in terms of a quantum transition readily available. Recall that our ansatz satisfied all equations of motion except those for (\ref{feom2}) and even then, in the limit $\kappa^2 \to 0$, only for those matrix elements that corresponded to the poles of the spin j fuzzy spheres. So if we were to consider the statement that the action is extremized with respect to variations in the $X_i$, we see that:

\begin{equation}
\label{sss}
\frac{\delta L}{\delta X_i} = 0 
\end{equation}

only for matrix elements that do not correspond to $m \to \pm j$ and for times such that $t \neq t_f$. Let us consider the action evaluated for trajectories close to our ansatz:

\begin{equation}
\label{sans}
S = S_{ansatz} + \int dt~ Tr \frac{\delta L}{\delta X_i}\delta X_i 
\end{equation}

The action for our ansatz will scale with the number of independent degrees of freedom, which for our ansatz goes as $3jN$, where j is the dimension of the block partitions and N (which will tend to infinity) corresponds to the number of these blocks. However by (\ref{sss}) and the discussion above, although the first order variation vanishes for a great deal of matrix elements, the elements for which it does not vanish go as $2N$. We arrive at this from the realisation that we only violated the equations of motion at the boundaries of the block representations, for times close to $t_f$, and hence scales as twice the number of blocks (the factor 2 comes from accounting for the analogous equation for $X_-$). So if we were to consider the transition amplitude given by:

\begin{equation}
\label{path}
\int \mathcal{D} X e^{-\int^{s}_{c} dt~ S[X]}
\end{equation}

where the limits on the integration correspond to path integrating over all trajectories that begin on the cylinder solution and end on the infinite collection of spin j spheres. One could have one of two situations to contend with-- either a true classical solution exists which our ansatz fails to capture, or such a solution does not exist. If the prior case is true, then we have nothing to prove. However if such a solution does not exist, then consider the contribution to the path integral coming from the trajectories close to that described by our ansatz-- one can conclude from the considerations above that:

\begin{equation}
\label{jklm}
S_{ansatz} \sim O(3jN) ~\gg~ \frac{\delta S}{\delta X}|_{ansatz} \sim O(2N)
\end{equation}

One recognizes in these inequalities, the conditions for the path integral to be dominated by this approximately stationary point. Hence in the absence of a truly classical path interpolating between the fuzzy cylinder and a collection of fuzzy spheres, our ansatz will be among the dominant contributions to this path integral. That is, one can also interpret our ansatz as the leading contribution to the quantum transition amplitude between the two static solutions.

\section{Closing Remarks}

Although topology changing dynamics has been a long suspected component of matrix mechanics, an explicit demonstration has till now proven to be somewhat elusive. Inspite of its apprerance, the existence of such a demonstration is not a trivial problem to be left forgranted, as finding such a solution will answer the question of whether or not such topology change is a classical or quantum phenomenon. As we have hopefuly demonstrated, using matrix models for their ability to model the dynamics of non-commuting spaces has shown topology change to be primarily a classical phenomenon in the large N limit. One should interpret classical in this sense with care, as we are already implicitly working in a highly 'quantum mechanical' framework in proposing spacetime non-commutativity. The more apropriate statement being that there is no need to `third quantize' our model to see topology change. There are many avenues for future investigations in light of these results. We feel that matrix models might be a highly tractable bootstrap into a viable theory of spacetime at the Planck length, and perhaps will imply what one already suspects is the case from the collective experience of string theory-- that geometry is a concept derived from the physics of our degrees of freedom.

\section{Acknowledgements}

The author wishes to thank the Theory group at the Tata Institute for Fundamental Research in Mumbai, for their hospitality and support throughout the period in which this work was undertaken. During rather difficult personal circumstances, the ever humourous presence of Rudra Pratap Jena and the patient support of Sandip Trivedi made it easier for the author to absorb himself in physics. The author wishes to thank Gautam Mandal and K.P. Yogendran for their ideas and guidance, Robert Brandenberger for continued encouragemnet, and A.P. Balachandran for comments on the manuscript. Thanks must also go to Bernard Wolfsdorf, Angela Wong and Naveen Rahman for continued help throughout the period in which this work was undertaken.

\end{document}